# Robust magnetism against pressure in non-superconducting samples prepared from lutetium foil and $H_2/N_2$ gas mixture

Jing Guo[1]*, Shu Cai[1,2]*, Dong Wang[2], Haiyun Shu[2], Liuxiang Yang[2], Pengyu Wang[1,3], Wentao Wang[1], Huanfang Tian[1], Huaixin Yang[1], Yazhou Zhou[1], Jinyu Zhao[1,3], Jinyu Han[1,3], Jianqi Li[1], Qi Wu[1], Yang Ding[2], Wenge Yang[2], Tao Xiang[1,3], Ho-kwang Mao[2] and Liling Sun[1,2,3]†

[1] *Institute of Physics, Chinese Academy of Sciences, Beijing 100190, China*
[2] *Center for High Pressure Science & Technology Advanced Research, 100094 Beijing, China*
[3] *University of Chinese Academy of Sciences, Beijing 100190, China*

Recently, the claim of "near-ambient superconductivity"[1] in a N-doped lutetium hydride attracted enormous following-up investigations in the community of condensed matter physics and material sciences[2-11]. But quite soon, the experimental results from different groups indicate consistently that no evidence of near-ambient superconductivity is found in the samples synthesized by the same method as the reported one [2], or by the other alternative methods [3-11]. From our extended high-pressure heat capacity and magnetic susceptibility measurements on the samples prepared with the lutetium foil and $H_2/N_2$ gas mixture, we report the finding of a magnetic transition at the temperature about 56 K. Our results show that this magnetic phase is robust against pressure up to 4.3 GPa, which covers the critical pressure of boosting the claimed near room temperature superconductivity[1].

The heat capacity is a straightforward measure to the magnetic transition or the bulk nature of superconductivity of materials, thus it has been widely used in the studies of the magnetism in electron correlated systems [13-18]. However, precise measurement in high-pressure environment is very challenge due to the tiny amount of sample that can only produce a weak signal. By using our newly developed state-of-the-art technique - quasi-hydrostatic-pressure heat capacity measurements in a diamond anvil cell, we performed the heat capacity measurements on the two types of lutetium-hydrogen-nitrogen samples: one is prepared from the lutetium foil and $H_2/N_2$ gas mixture (mole ratio 99:1) followed by heating to 65° C for 24 hours, the same procedure as that described by Dasenbrock-Gammon et al [1] (we define this sample as the "Lu-N-H-65°C" in short ); and the other is also prepared using the same gas mixture of hydrogen and nitrogen, but followed by heating to 1800° C with a laser heating technique (we define this sample as "Lu-N-H-1800° C" in short). Figure 1a and 1b show the experimental setup for the heat capacity measurements. This arrangement allows us to retrieve the sample's the heat capacity value by converting the heater's modest temperature oscillation into an *ac* voltage signal. The experimental details can be found in the Supplementary Information (SI).

Before the high-pressure measurements, we first characterize the chemical composition of the investigated samples using the spatially resolved electron energy-loss (EELS) spectroscopy equipped in a scanning transmission electron microscope (STEM). As shown in Fig. 1c and 1d, the EELS spectra of the sample Lu-N-H-65°C were measured within two energy ranges (1225 eV-2050 eV and 375 eV-810 eV), in which the green lines in Fig.1c and Fig. 1d represent the spectra after background subtraction (black lines are the raw data, while the red lines are the background). The clear signals above ~1585 eV and ~535 eV are from Lu and O elements, respectively. However, no signal around 400 eV for nitrogen was detected, suggesting that no trace of nitrogen is incorporated in our sample, consistent with the energy dispersive spectroscopy (EDS) results [2].

Next, we performed high-pressure heat capacity measurements on the sample Lu-N-H-65 °C. Figure 2(a)-2(d) show the temperature versus $C/T$ measurements in the pressure range 1.3 GPa - 4.3 GPa and temperature down to 4 K. One can see that the $C/T$ data exhibit considerable dispersion at the temperature higher than ~80 K, which may be caused by the weak signal of the small sample (90 μm × 70 μm × 20 μm) and the substantial contribution of phonons in this regime. Nevertheless, our results show no distinguishable change in the high temperature range from 300 K down to 80 K, on the contrary to the report by Dasenbrock-Gammon *et al*[1] - a distinctive discontinuity in the heat capacity measured from the sample with the similar size to ours in the pressure range 1 - 2 GPa at near room temperature (such a discontinuity was claimed to be related to a superconducting transition[1]). The absence of the near-room temperature superconductivity in the compressed Lu-H-N sample investigated by our heat capacity measurements is consistent with the results obtained from our resistance and magnetic susceptibility measurements[2]. However, intriguingly we found an anomaly in the plot of $C/T$ versus temperature at ~55 K (Fig.2a-2d).

To confirm the anomaly detected by our heat capacity measurements, we conducted high-pressure *dc* magnetic susceptibility measurements on the same sample. As shown in Fig.3a-3c, the plots of magnetization versus temperature display a drop at ~ 57 K upon cooling for pressures ranging from 0.8 GPa to 3.3 GPa. With increasing pressure, the drop is getting more pronounced. The observations of the heat capacity anomaly and the magnetic susceptibility drop suggest a magnetic transition. We applied magnetic field on the sample subjected to 3.3 GPa and found that the transition temperature shifts to lower temperature initially when the field is enhanced from 20 Oe to 200 Oe, and then remains almost unchanged with further increment of magnetic field up to 4000 Oe (Fig.3d). The robust magnetism against field, together with our results of the resistance measurement that shows a continuous decrease upon cooling across this temperature[2], reveals that the 'transition' at ~ 56 K (we take an average value obtained by heat capacity and magnetic susceptibility measurements (55+57)/2=56 K) should be associated with a magnetic transition, instead of a weak signal from superconducting transition.

The same magnetic transition was also observed in the sample-Lu-N-H-1800 °C in the pressure range 0.2 – 3.2 GPa (see SI). The transition occurs at ~ 56 K at 0.2 GPa and prevails to 57.8 K at 3.2 GPa, also exhibiting the robust magnetism as that observed from the sample Lu-H-N-65 °C. To justify the reliability and accuracy of our measurements, we performed heat capacity measurements, by employing the same experimental setup, on the $CaK(Fe_{0.96}Ni_{0.04})_4As_4$ superconductor that holds a spin vertex crystal state (a type of magnetic state) at 40 K and superconducting transition at

21 K at ambient pressure [19-22]. The results (see SI) demonstrate that the experimental setup for the high-pressure heat capacity is reliable, and the data detected from the Lu-H-N samples are believable.

We summarize our high-pressure results obtained from the heat capacity and magnetic susceptibility measurements on the two samples prepared at 65° C and 1800° C in the pressure-temperature phase diagram (Fig. 4). It is shown that their magnetic transition temperatures ($T_M$) vary within a narrow temperature regime in the pressure range investigated. The existence of the robust magnetic phase in these two distinct samples rules out the possibility that they host any superconducting phase, even in a minor amount, at the low temperature regime.

The same magnetic transition found in both samples with heat treatment at 65 °C and 1800 °C is reminiscent of what we observed in X-ray diffraction measurements, in which the same substance, $LuH_2$, presents in these samples. Moreover, the recent studies on $LuH_2$ revealed that it undergoes a magnetic transition at ~ 60 K [9], which is very close to the temperature of the magnetic phase suggested from the observation of heat capacity and magnetic measurements. As a result, we suggest that this magnetic phase should be originated from $LuH_2$ phase in our samples.

In summary, we performed systematic investigations with heat capacity and magnetic susceptibility measurements in a diamond anvil cell, on the two different heat-treated samples that are initially prepared with the same method as that reported in Ref.1 - using $99H_2/1N_2$ gas mixture and a lutetium foil as starting materials. Before the high-pressure experiments, we treated one of the samples with the same annealing method as reported by Dasenbrock-Gammon *et al*[1] (heating the samples at 65°C for 24 hours), and separately heating the other one to 1800° C for several minutes by the laser heating technique. Our high-pressure results measured from the heat capacity and magnetic susceptivity consistently indicate the existence of a magnetic phase, whose transition temperature is ~ 56 K, in these two types of samples. Upon increasing pressure up to 4.3 GPa, the magnetic transition temperature ($T_M$) of the two samples displays a small variation with pressure. The presence of the robust magnetic phase in the low temperature and the absence of superconductivity imply that this physical system may host some interesting physics but not superconductivity.

These authors with star (*) contributed equally to this work.
Correspondence and requests for materials should be addressed to L.S. (llsun@iphy.ac.cn)

**Acknowledgements**

This work was supported by the National Key Research and Development Program of China (Grant No. 2022YFA1403900 and 2021YFA1401800), the NSF of China (Grant Numbers Grants No. U2032214, 12104487, 12122414 and 12004419), and the Strategic Priority Research Program (B) of the Chinese Academy of Sciences (Grant No. XDB25000000). J. G. and S.C. are grateful for supports from the Youth Innovation Promotion Association of the CAS (2019008) and the China Postdoctoral Science Foundation (E0BK111).

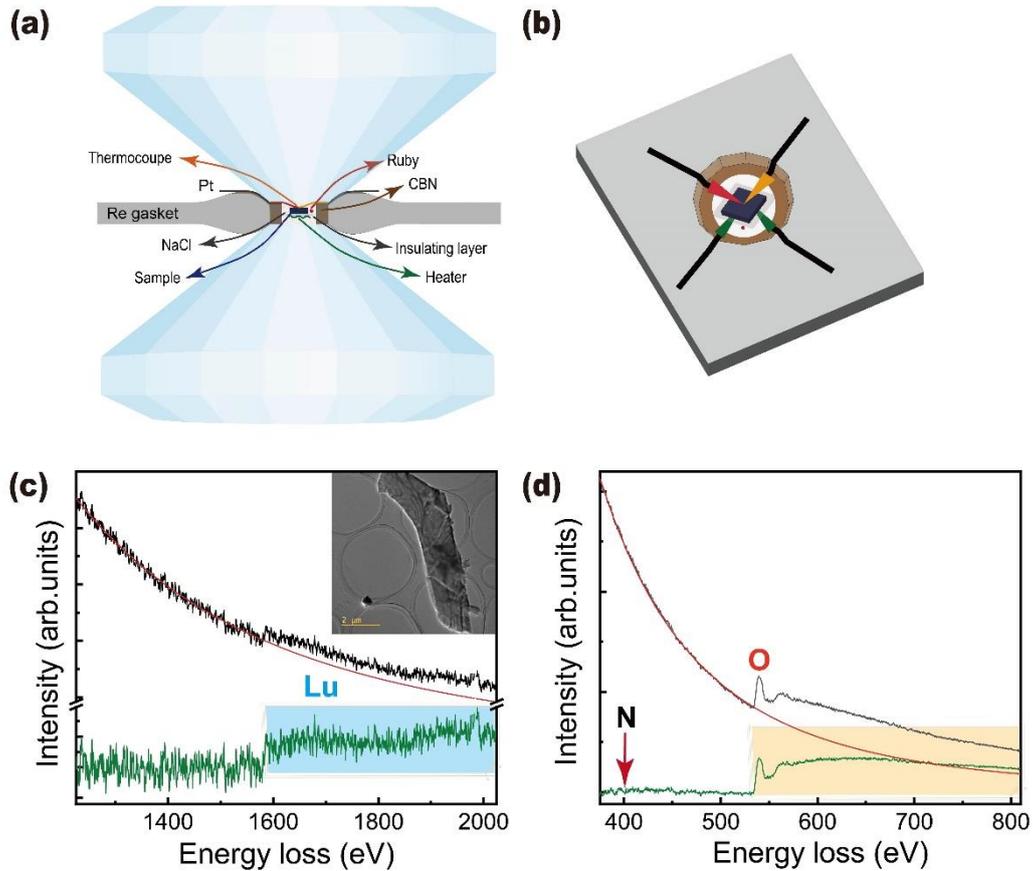

**Fig 1. Schematic illustration of the high-pressure heat capacity measurement configuration and characterization of the chemical composition for the Lu-H-N sample.** (a) Experimental arrangement for the high-pressure heat capacity measurement in a diamond anvils cell. (b) A top view of the experimental setup that displays the heater, thermocouple, insulating layer and Pt leads in the sample area. The thermocouple (see red and orange arrows) was positioned on the one side of the sample to measure the temperature oscillation, and a constantan heater (see green arrows) that was placed to the opposite side. (c)-(d) The EELS spectra of the sample Lu-N-H-65°C. The inset of the figure (c) displays the sample on the copper net.

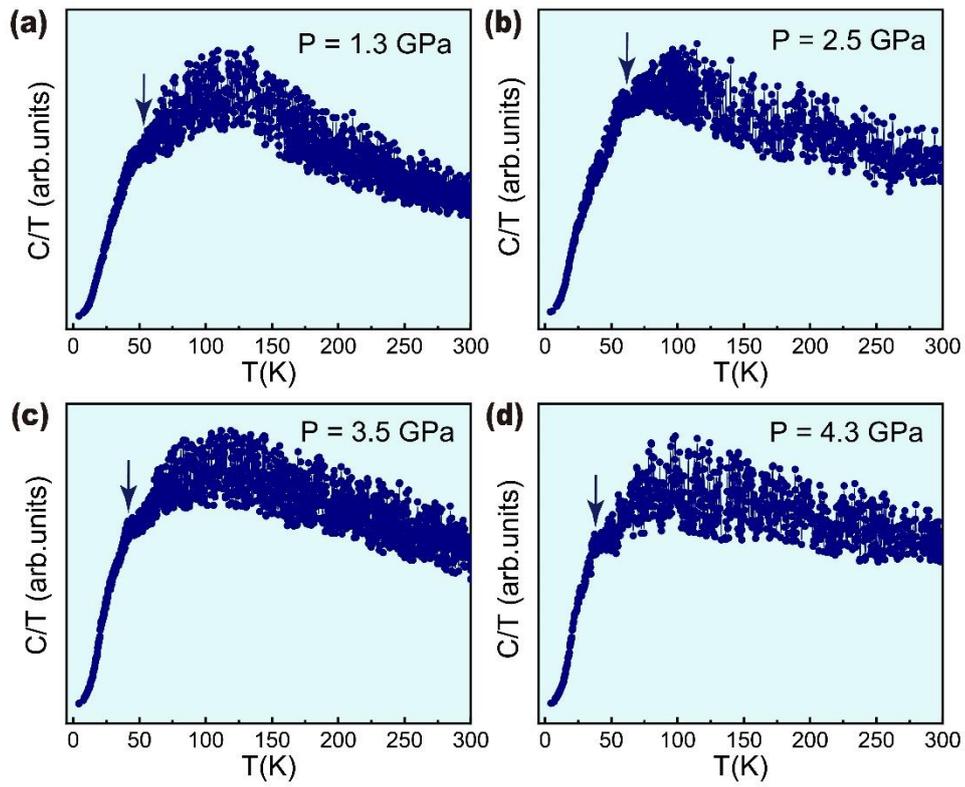

**Fig. 2 The results obtained from high-pressure heat capacity measurements on the sample Lu-N-H-65°C.** (a)-(d) Heat capacity (*C/T*) as a function of temperature in the pressure range 1.3-4.3 GPa and temperature range 4 -300 K.

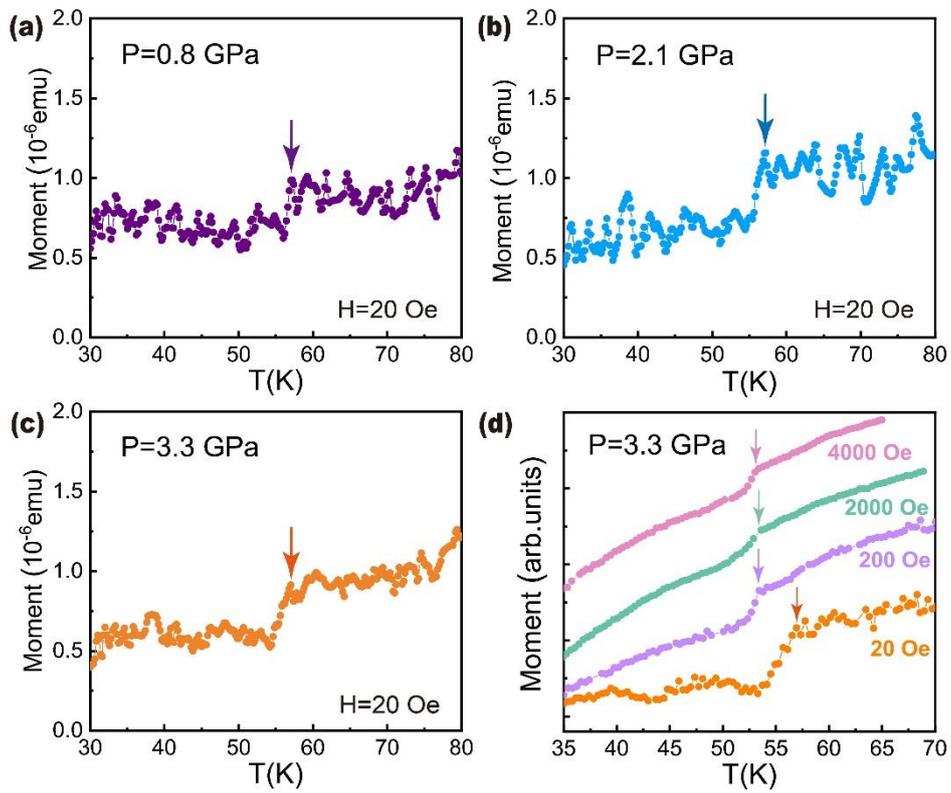

**Fig3. The results of high-pressure magnetic susceptibility measured at different pressures for the sample Lu-N-H-65°C.** (a)-(c) Magnetic susceptibility versus temperature measured in the pressure range 0.8-3.3 GPa and temperature down to 30 K. The data were obtained after the background subtraction. (d) The magnetic susceptibility of the sample subjected to 3.3 GPa under different magnetic fields.

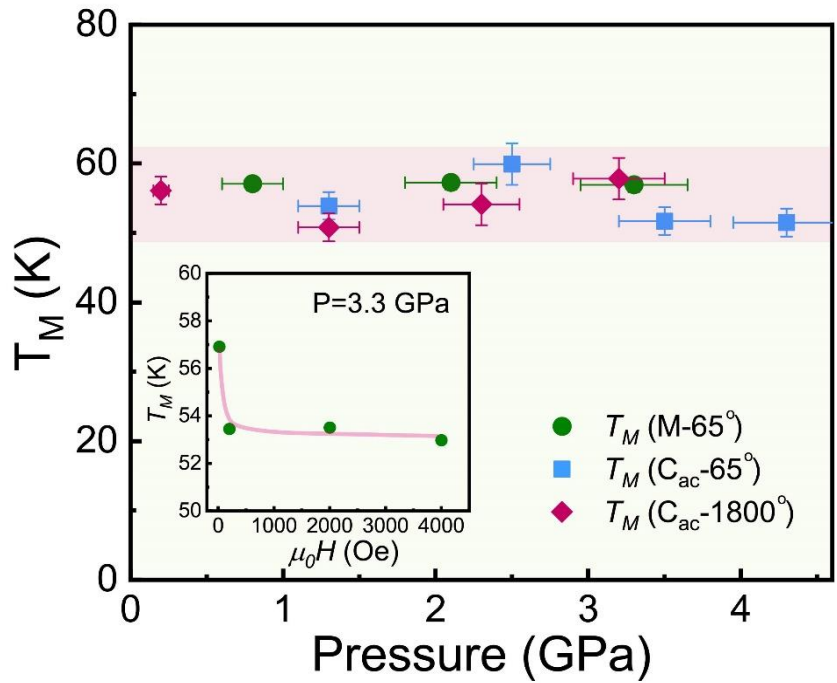

**Fig 4 Pressure-temperature phase diagram for the two compressed Lu-H-N samples by different heat treatments**. $T_M$ (M-65°C) and $T_M$ (Cac-65°C) represent the magnetic transition temperature detected by the measurements of magnetic susceptibility and the heat capacity on the sample Lu-N-H-65°C, respectively. $T_M$ (Cac-1800°C) stands for the magnetic transition temperature measured through the heat capacity measurements on the sample Lu-N-H-1800°C. The inset shows the magnetic transition temperature as a function of magnetic field for the sample-obtained-at 65°C at 3.3 GPa.